\documentclass[aps,prl,reprint,superscriptaddress,twocolumn,longbibliography]{revtex4-1}
\usepackage{amssymb}
\usepackage{amsmath}
\usepackage{graphicx}
\usepackage{epstopdf}
\usepackage{times}
\usepackage[colorlinks, linkcolor=blue, urlcolor=blue, anchorcolor=blue, citecolor=blue]{hyperref}

\begin{document}


\title{Observation of Non-Markovian Spin Dynamics in a Jaynes-Cummings-Hubbard Model using a Trapped-Ion Quantum Simulator}
\author{B.-W. Li}
\thanks{These authors contribute equally to this work}%
\affiliation{Center for Quantum Information, Institute for Interdisciplinary Information Sciences, Tsinghua University, Beijing 100084, P. R. China}
\author{Q.-X. Mei}
\thanks{These authors contribute equally to this work}%
\affiliation{Center for Quantum Information, Institute for Interdisciplinary Information Sciences, Tsinghua University, Beijing 100084, P. R. China}
\author{Y.-K. Wu}
\thanks{These authors contribute equally to this work}%
\affiliation{Center for Quantum Information, Institute for Interdisciplinary Information Sciences, Tsinghua University, Beijing 100084, P. R. China}
\author{M.-L. Cai}
\affiliation{Center for Quantum Information, Institute for Interdisciplinary Information Sciences, Tsinghua University, Beijing 100084, P. R. China}
\affiliation{HYQ Co., Ltd., Beijing, 100176, P. R. China}
\author{Y. Wang}
\affiliation{Center for Quantum Information, Institute for Interdisciplinary Information Sciences, Tsinghua University, Beijing 100084, P. R. China}
\author{L. Yao}
\affiliation{Center for Quantum Information, Institute for Interdisciplinary Information Sciences, Tsinghua University, Beijing 100084, P. R. China}
\affiliation{HYQ Co., Ltd., Beijing, 100176, P. R. China}
\author{Z.-C. Zhou}
\affiliation{Center for Quantum Information, Institute for Interdisciplinary Information Sciences, Tsinghua University, Beijing 100084, P. R. China}
\author{L.-M. Duan}
\email{lmduan@tsinghua.edu.cn}
\affiliation{Center for Quantum Information, Institute for Interdisciplinary Information Sciences, Tsinghua University, Beijing 100084, P. R. China}


\begin{abstract}
Jaynes-Cummings-Hubbard (JCH) model is a fundamental many-body model for light-matter interaction. As a leading platform for quantum simulation, the trapped ion system has realized the JCH model for two to three ions. Here we report the quantum simulation of the JCH model using up to 32 ions. We verify the simulation results even for large ion numbers by engineering low excitations and thus low effective dimensions; then we extend to 32 excitations for an effective dimension of $77$ qubits, which is difficult for classical computers. By regarding the phonon modes as baths, we explore Markovian or non-Markovian spin dynamics in different parameter regimes of the JCH model, similar to quantum emitters in a structured photonic environment. We further examine the dependence of the non-Markovian dynamics on the effective Hilbert space dimension. Our work demonstrates the trapped ion system as a powerful quantum simulator for many-body physics and open quantum systems.
\end{abstract}

\maketitle

As the size and the controllability of the available quantum information processors advance \cite{doi:10.1063/1.5088164,huang2020superconducting}, quantum simulation \cite{feynman2018simulating,georgescu2014quantum,cirac2012goals} has become a promising and convenient approach to understanding many-body quantum dynamics that are challenging for classical computers, for which various approximations have to be carefully designed and implemented due to the well-known ``curse of dimensionality''. Ion trap, one of the leading platforms for quantum information processing \cite{doi:10.1063/1.5088164}, has been widely applied in the quantum simulation of many-body spin models with long-range Ising or Heisenberg-type interactions \cite{PhysRevLett.92.207901,blatt2012quantum,monroe2021programmable} mediated by the spatial oscillation of the ions. Record-breaking experiments have simulated quantum dynamics of up to 53 spins \cite{zhang2017observation}, and properties such as phase transitions \cite{friedenauer2008simulating,islam2011onset,zhang2017observation,PhysRevLett.119.080501}, frustration \cite{kim2010quantum,islam2013emergence}, information propagation \cite{richerme2014non,jurcevic2014quasiparticle,PhysRevLett.124.240505}, localization \cite{smith2016many,brydges2019probing,PhysRevLett.122.050501,morong2021observation} and Floquet dynamics \cite{zhang2017time,kyprianidis2021prethermal} have been examined. In these experiments, the quantized oscillation modes, a.k.a. phonon modes, are only virtually excited through off-resonant laser driving. On the other hand, stronger driving close to phonon sidebands can explicitly excite the phonon states and interact them with the spins. Such schemes have found broad applications in quantum information processing \cite{Cirac1995,sorensen2000entanglement,Milburn2000MSgate,zhu2006trapped,leibfried2003experimental}, bosonic state engineering \cite{PhysRevLett.90.037902,um2016phonon,PhysRevLett.119.033602} and quantum transport \cite{PhysRevLett.111.040601,Ramm_2014,abdelrahman2017local,PhysRevX.8.011038,PhysRevLett.124.200501}. Moreover, the inclusion of the phonon degrees of freedom opens up an avenue toward the richer phenomena in the spin-boson hybrid systems such as the quantum Rabi model \cite{forn-diaz2019ultrastrong,lv2018quantum,cai2021observation} for a single ion and the Hubbard-like models \cite{greentree2006quantum,angelakis2007photonblockadeinduced,PhysRevLett.109.053601} in the multi-ion cases.

Two prototypical many-body models with spin-boson interactions are the Jaynes-Cummings-Hubbard (JCH) model \cite{greentree2006quantum,angelakis2007photonblockadeinduced,hartmann2006strongly,rossini2007mottinsulating,PhysRevA.77.033801,makin2008quantum,koch2009superfluid,schmidt2009strong,ivanov2009simulation} and the Rabi-Hubbard (RH) model \cite{PhysRevLett.109.053601,hwang2013largescale,zhu2013dispersive,schiro2013quantum,flottat2016quantum}, both of which originate from cavity quantum electrodynamics systems but are well-suited for the trapped ions owing to the strong and controllable spin-phonon coupling. The embracement of the spin and phonon states significantly increases the dimension of the effective Hilbert space, making these problems even more challenging for classical computers. Recently, equilibrium and dynamical properties of the RH model has been studied for up to 16 ions, which amounts to the complexity of about 57 qubits \cite{mei2021experimental}. In comparison, the JCH model possesses an additional $U(1)$ symmetry and thus demonstrates essentially different properties: The ground state phase diagram of the JCH model now displays a multicritical point \cite{koch2009superfluid,schmidt2009strong} similar to the Bose-Hubbard model \cite{fisher1989boson,jaksch1998cold,bloch2008manybody}, as opposed to the Ising universality class of the RH model \cite{schiro2013quantum,flottat2016quantum}; the effective Hilbert space dimension is now governed by the conserved excitation number of the system, allowing a well-regulated study of the many-body dynamics versus the system dimension from a polynomial to an exponential scaling with the ion number. This also provides a natural test bed to directly verify the simulated Hamiltonian for large ion numbers, rather than the exponential cost for general systems \cite{PhysRevLett.124.010504,PRXQuantum.2.010102,zhu2021cross} or to extrapolate from smaller systems.  Furthermore, when regarding the phonon modes as an environment, the model resembles quantum emitters in a structured photonic background like a photonic crystal \cite{PhysRevA.50.1764,thompson2013coupling,goban2014atom,PhysRevLett.119.143602} where peculiar phenomena can emerge such as non-Markovian dynamics \cite{PhysRevLett.106.233601,PhysRevLett.108.043603,PhysRevA.96.043811,PhysRevA.99.032101,ferreira2021collapse}, collective radiation \cite{arjan2013photon,PhysRevLett.115.063601,PhysRevLett.117.133603,PhysRevA.96.043811} and the dissipative generation of entanglement \cite{Gonzalez_Ballestero_2013,PhysRevLett.99.160502,PhysRevLett.108.160402,PhysRevLett.115.163603}. Previously, the JCH model has been implemented in ion trap in small scales using two \cite{toyoda2013experimental,ohira2021blockade} or three ions \cite{debnath2018observation} for observing the hopping and blockade of phonons and the signature of quantum phase transition. However, many of the aforementioned dynamical properties require large system sizes and remain to be demonstrated in the experiment.

In this letter, we report the quantum simulation of a JCH model with long-range interaction using a trapped chain of up to 32 ions. The successful simulation of the JCH Hamiltonian is verified directly for large ion numbers by engineering low total excitation number of the system. Then we demonstrate the change from the Markovian to non-Markovian dynamics by tuning the frequency of the spins into different locations of the phonon spectrum. We further adjust the effective dimension of the system via the ion number and the excitation number, and observe that the non-Markovian dynamics persists for large systems. With up to 32 excitations in 32 ions, an effective Hilbert space dimension above $2^{77}$ is achieved, which is challenging for existing supercomputers. Our work showcases the trapped ion system as a powerful quantum simulator of spin-boson coupled systems and open quantum systems with a bosonic environment.

\begin{figure}[!tbp]
	\centering
	\includegraphics[width=0.9\linewidth]{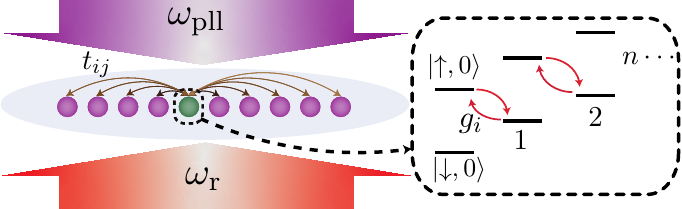}
	\caption {Experimental scheme. Two counter-propagating global Raman laser beams are applied perpendicular to a chain of 2 to 32 ions to generate a Jaynes-Cummings-Hubbard (JCH) model Hamiltonian. The beat frequency of the Raman laser beams is tuned close the red motional sideband, and is stabilized by a phase-locked loop (PLL). The internal state of each ion is coupled with its local oscillation by the Raman laser beams as a Jaynes-Cummings model, and these local oscillation modes of individual ions are further coupled by their Coulomb interaction. \label{fig1}}
\end{figure}

Our experimental setup is shown schematically in Fig.~\ref{fig1} using a chain of ${}^{171}\mathrm{Yb}^{+}$ ions in a linear Paul trap. The internal electronic levels $|\downarrow\rangle\equiv |\mathrm{S}_{1/2}, F = 0, m_F = 0\rangle$ and $|\uparrow\rangle\equiv|\mathrm{S}_{1/2}, F = 1, m_F = 0\rangle$ of each ion encode an individual spin. The local phonon modes of the ion chain naturally possess long-range hopping terms due to the Coulomb interaction. We further turn on the on-site Jaynes-Cummings interaction via a pair of counter-propagating $355\,$nm global Raman beams whose beat frequency is tuned close to the red motional sideband. Moving into an interaction picture, the Hamiltonian of the system is governed by a JCH model
\begin{align}
H =& \sum_i \left[\frac{\Delta}{2} \sigma_z^i + \omega_i a_i^\dagger a_i + g_i (\sigma_+^i a_i + a_i^\dagger \sigma_-^i) \right] \nonumber\\
& +\sum_{i<j} t_{ij}(a_i^\dag a_j +a_j^\dagger a_i),
\end{align}
where $\Delta$ is the spin frequency set by the Raman laser detuning from the motional sideband, $\omega_i$ the local phonon frequency in the interaction picture, $g_i$ the spin-phonon coupling on the $i$th ion, and $t_{ij}$ the phonon hopping rates. Since $g_i$ is not uniform due to the finite laser beam width, we describe it by a maximal coupling $g$ for the central ion and the others can be deduced from the Gaussian beam profile of the laser. More details about the derivation of the Hamiltonian and the calibration of the model parameters can be found in Supplementary Materials \cite{supplementary}. Note that $\Delta$ and $\omega_i$ can be shifted by a same constant without affecting any relevant dynamics owing to the $U(1)$ symmetry of the Hamiltonian. This symmetry also results in the conservation of total spin and phonon excitations in the system, which is in stark contrast to the RH model \cite{mei2021experimental}.

\begin{figure}[!tbp]
	\centering
	\includegraphics[width=0.9\linewidth]{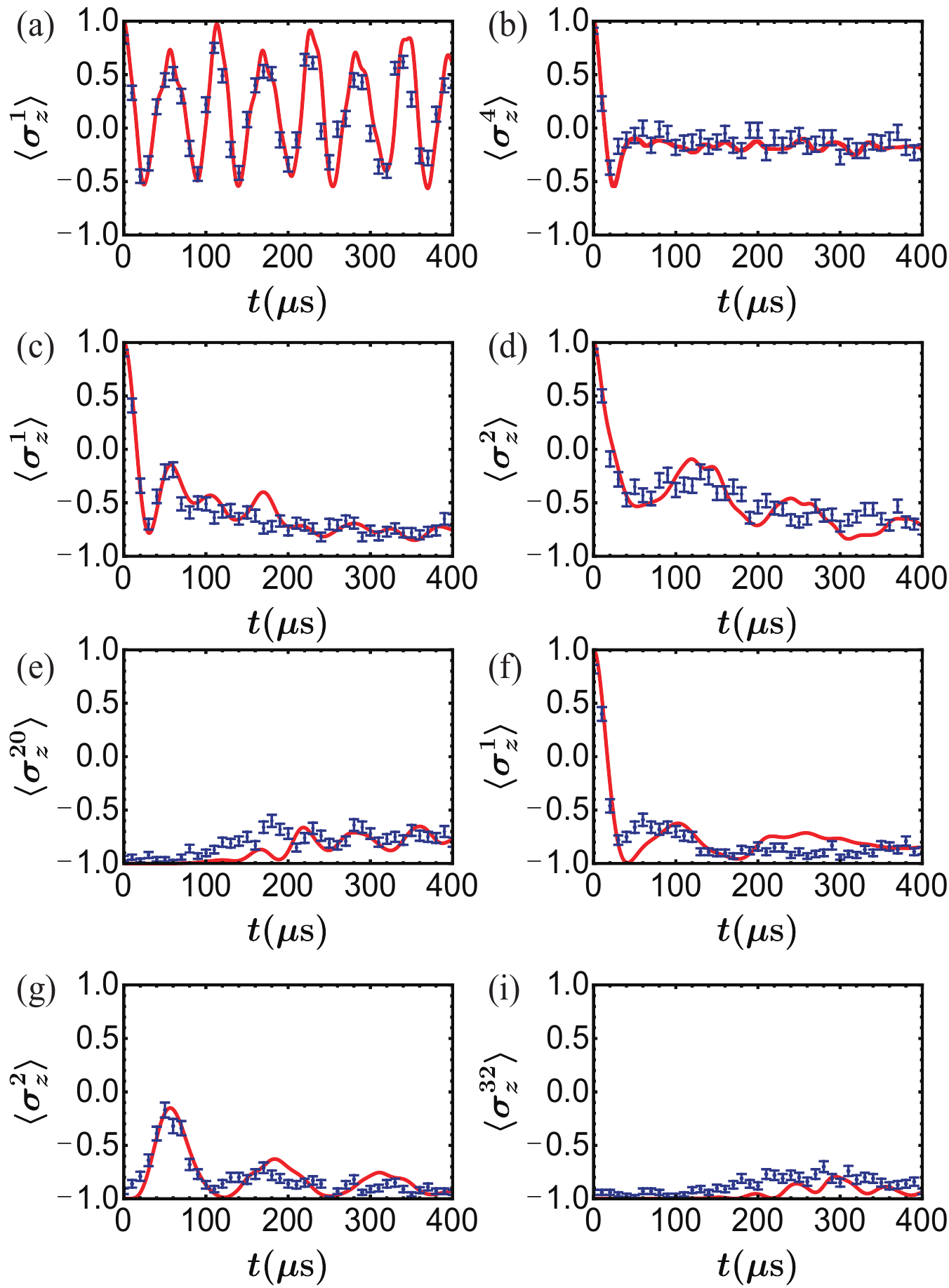}
	\caption {Comparison between experimental spin dynamics and theoretical prediction for small Hilbert space dimensions. Blue dots (with error bars representing one standard deviation) are experimentally measured $\langle\sigma_z^i(t)\rangle$ for individual ions, and the red curves are the corresponding theoretical results by exact diagonalization with no fitting parameters. (a) Two ions initialized in $|\uparrow,0\rangle^{\otimes 2}$ (two total excitations) under $\Delta = -2\pi\times 60\,$kHz and $g = 2\pi\times 11.6\,$kHz. The evolutions of the two ions are symmetric so only one is plotted. (b) Similar plot for a central ion (ion 4, labelled by 1 to $N$ from left to right) in an $N=8$ chain under $\Delta = -2\pi\times 60\,$kHz and $g = 2\pi\times 11.5\,$kHz with the initial state $|\uparrow,0\rangle^{\otimes 8}$ (8 excitations). (c-e) $N=20$ ions with two total excitations initialized in $|\uparrow,0\rangle^{\otimes 2}\otimes|\downarrow,0\rangle^{\otimes 18}$ under $\Delta = -2\pi\times 5\,$kHz and $g = 2\pi\times 10\,$kHz. (c) Ion 1 and (d) ion 2 are the two ions being excited on the left of the chain, while (e) ion 20 is the ion on the right. (f-h) $N=32$ ions with one total excitation initialized in $|\uparrow,0\rangle\otimes|\downarrow,0\rangle^{\otimes 31}$ under $\Delta = -2\pi\times 5\,$kHz and $g = 2\pi\times 11.6\,$kHz. (f) is for the excited ion 1 on the left, (g) for its neighbor ion 2 and (h) for the ion 32 on the other end of the chain. \label{fig2}}
\end{figure}

\begin{figure*}[!tbp]
	\centering
	\includegraphics[width=0.9\linewidth]{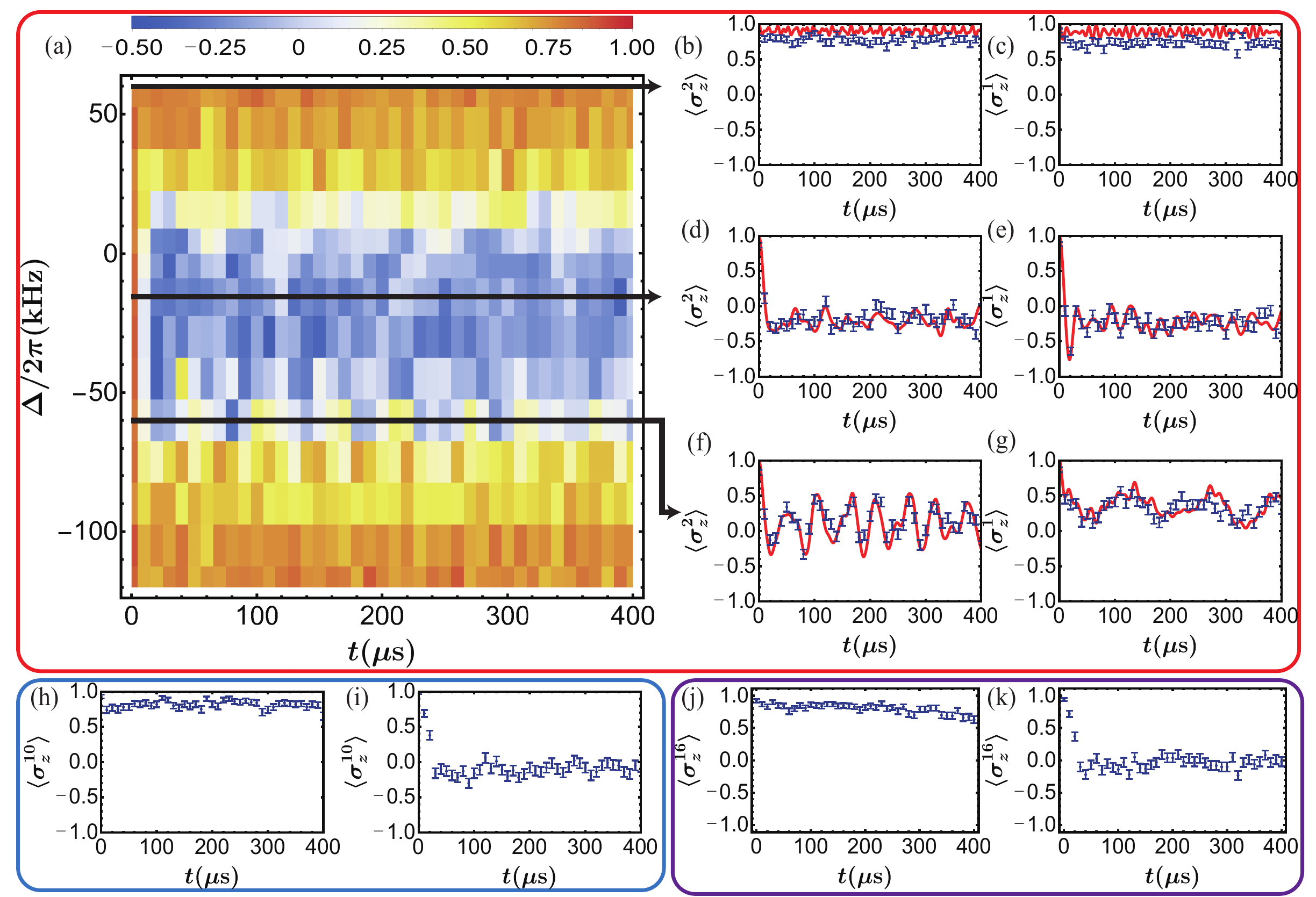}
	\caption {Markovian and non-Markovian spin dynamics in different regimes of the JCH model. We initialize the system at $|\uparrow,0\rangle^{\otimes N}$ with $M=N$ total excitations. We fix $g=2\pi\times 12.9\,$kHz and tune $\Delta$ to observe different spin dynamics $\langle\sigma_z^i(t)\rangle$. (a) Experimental results for $N=4$ ions with $\Delta$ ranging from $-2\pi\times 120\,$kHz to $2\pi\times 60\,$kHz. The response strongly depends on the location of $\Delta$ with respect to the phonon band from $-2\pi\times 52\,$kHz to zero. (b-g) Dynamics in different parameter regimes in (a) as indicated by black horizontal arrows for the central (left panel) and edge (right panel) ions. Red curves are the theoretical results from exact diagonalization. (h, i) Measured spin dynamics for a central ion of an $N=20$ chain with $\Delta = 2\pi\times 30\,$kHz (outside band) and $-2\pi\times\,5$kHz (near band edge), respectively. (j, k) Similar plots for an $N=32$ chain with $\Delta = 2\pi\times30\,$kHz and $-2\pi\times5\,$kHz, respectively. \label{fig3}}
\end{figure*}

First we verify the successful quantum simulation of the JCH model by measuring the spin dynamics and comparing with the theoretical predictions in small system sizes. For $N$ ions and $M$ total excitations, the effective dimension of the Hilbert space is given by
\begin{equation}
D=\sum_{k=0}^{\min(N,\,M)} C(N,\,k)\times C(N+M-k-1,\, N-1),
\end{equation}
where $C(n,m)\equiv n!/[m!(n-m)!]$ is the combination number to choose $m$ items from $n$ elements. Although $D$ generally increases rapidly with $N$ and $M$ (for $N=M=32$ we have $D>2^{77}$), it turns out that even under large ion number $N=32$, the classical simulation is still feasible for small values of $M$. This allows us to directly verify the correctness of the quantum simulation results at large system sizes rather than to extrapolate from smaller systems. In Fig.~2 we plot the measured spin dynamics for $N=2$ and $N=8$ ions with $M=N$ total excitations, and for $N=20$ and $N=32$ ions with $M=2$ and $M=1$ total excitations, respectively. To create $M=N$ total excitations, we initialize the system in $|\psi_0\rangle=|\downarrow,\,0\rangle^{\otimes N}$ by sideband cooling and optical pumping \cite{leibfried2003quantum,supplementary}, and then we apply a global microwave $\pi$ pulse to get $|\uparrow,\,0\rangle^{\otimes N}$. As for the $M=1$ or $M=2$ cases, after initializing $|\psi_0\rangle$, we use a combination of global microwave pulses and a focused $355\,$nm Raman beam to flip the target ions (which we choose as ions on the edges because they have larger inter-ion spacings and thus less crosstalk errors) into $|\uparrow\rangle$ \cite{supplementary}. After preparing an initial state with the desired total excitation number, we turn on the JCH Hamiltonian and measure $\langle\sigma_z^i(t)\rangle$ for individual spins. As we can see, for various system sizes, excitation numbers as well as ion locations, the measured spin dynamics agrees well with the theoretical results with no free parameters (all parameters are calibrated in advance as described in Supplementary Materials \cite{supplementary}). There is a small systematic discrepancy for $N=20$ and $N=32$ where we prepare one or two spin excitations on the one end of the chain while the curve for the ion on the opposite end rises earlier than the theoretical prediction. This is mainly caused by the imperfect sideband cooling such that with small probability there are additional phonon excitations in the system that can quickly convert into the spin excitation of the distant ions \cite{supplementary}.

After demonstrating the successful quantum simulation of the JCH model, now we regard the phonon modes as a structured bosonic environment and examine the non-Markovian dynamics of the spins. This problem is similar to quantum emitters in a photonic crystal \cite{PhysRevA.50.1764,thompson2013coupling,goban2014atom,PhysRevLett.119.143602}, and it has been predicted \cite{PhysRevLett.119.143602,PhysRevA.96.043811} and demonstrated \cite{ferreira2021collapse} that by placing the spins at different locations of the bosonic spectrum, strikingly different behavior can occur.
In Fig.~\ref{fig3} we measure the spin dynamics for $N=4,\,20,\,32$ and $M=N$ from the initial state $|\uparrow,0\rangle^{\otimes N}$. For $N=4$ ions the frequencies of the collective phonon modes (note that they are different from the local phonon frequencies $\omega_i$) are distributed from $-2\pi\times 52\,$kHz to 0 in the interaction picture \cite{supplementary}. As we can see in Fig.~\ref{fig3}(a), when tuning $\Delta$ within this phonon band, we observe significant decay in the individual $\langle \sigma_z^i(t)\rangle$ away from their initial values of one, while outside the phonon band the response is much weaker. We further plot the dynamics for typical $\Delta$ in Fig.~\ref{fig3}(b-g) together with the theoretical predictions under the same parameters. Far outside the phonon band (b, c, $\Delta=2\pi\times 60\,$kHz), there is almost no decay in $\langle \sigma_z^i(t)\rangle$ due to the large detuning between the spins and phonons; inside the band (d, e, $\Delta=-2\pi\times 15\,$kHz) a fast decay is observed (predicted to be exponential in the continuum limit \cite{PhysRevLett.119.143602,PhysRevA.96.043811}) with small oscillations; and near the edge of the band (f, g, $\Delta=-2\pi\times 60\,$kHz) there is non-Markovian dynamics of both decay and oscillation, resulting from a mixture of the dynamics inside and outside the phonon band. We observe similar behavior for larger ion numbers $N=20$ (h, i) and $N=32$ (j, k) as well: outside the phonon band (h, j, $\Delta=2\pi\times 30\,$kHz) the spin population evolves slowly; while near the band edge we get non-Markovian dynamics of fast decay together with long-term oscillation, namely collapse and revival in the spin population.

\begin{figure}[!tbp]
	\centering
	\includegraphics[width=0.9\linewidth]{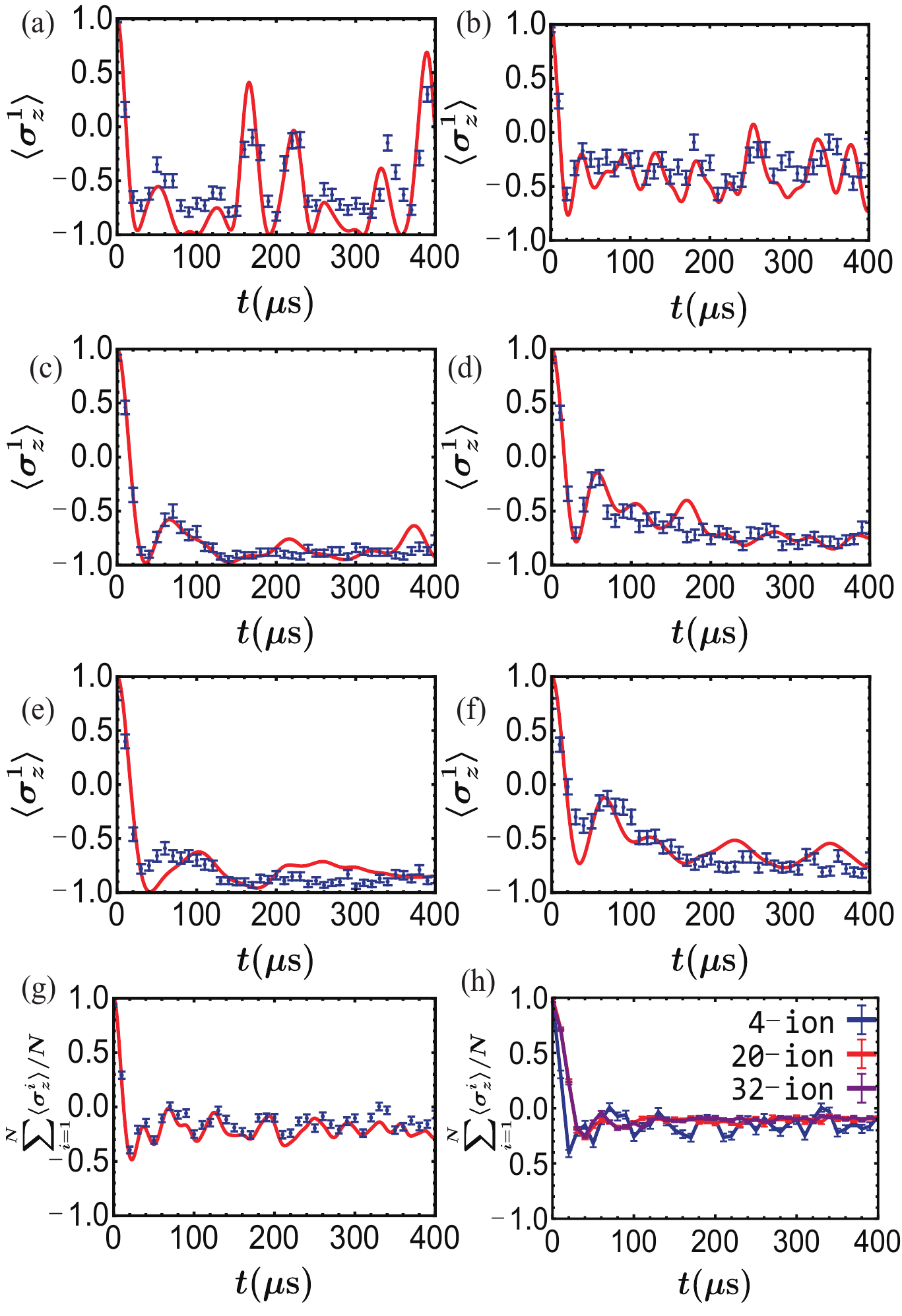}
	\caption {Collapse and revival in spin dynamics under various Hilbert space dimensions. (a, b) Experimental data (blue dots with error bars for one standard deviation) and theoretical results (red curves) for an initially excited ion in an $N=4$ chain with (a) one and (b) two total excitations, respectively. Here we choose $\Delta = -2\pi\times 10\,$kHz and $g = 2\pi\times 11.5\,$kHz. (c-f) Similar plots for (c, d) $N=20$ ions under $\Delta = -2\pi\times 5\,$kHz and $g = 2\pi\times 10\,$kHz, and (e, f) $N=32$ ions under $\Delta = -2\pi\times 5\,$kHz and $g = 2\pi\times 11.6\,$kHz. The system has (c, e) one or (d, f) two total excitations. (g) The average spin dynamics $\sum_i \langle\sigma_z^i(t)\rangle / N$ under the parameters of (a, b) with all the spins initialized in $|\uparrow\rangle$. (h) The average spin dynamics for $N = 4$, $20$, $32$ ions with $M=N$ total excitations. The $N=4$ case is the same as (g). For $N=20$ and $N=32$ ions with $N$ excitations, numerical calculation using exact diagonalization is intractable. Hence in this case the theoretical curves are not plotted, and the data points are directly connected to guide the eyes. \label{fig4}}
\end{figure}

Intuitively, one would expect a monotonic decrease in the revival signal as the dimension of the effective Hilbert space increases. However, the non-Markovian dynamics is known to persist even for a continuum of the environment \cite{PhysRevA.96.043811,ferreira2021collapse}. To resolve this inconsistency, in Fig.~\ref{fig4} we systematically examine this dependence on the system dimension. We plot the spin dynamics for $N=4$ (a, b), $N=20$ (c, d) and $N=32$ (e, f) ions with $M=1$ (left) and $M=2$ (right) total excitations. The change is most significant when increasing from $N=4$, $M=1$ to $N=4$, $M=2$ or to $N=20$, $M=1$, while for larger system sizes the curves are similar and there can even be additional revival signals owing to the added excitation number at $M=2$, which are also confirmed by the theoretical calculation as the red curves in these plots. This can be explained by the localization of the excitations near the individual spins when tuned close to the band edge \cite{PhysRevA.96.043811} such that the relevant phonon environment ceases to further scale up with the system sizes and thus the dynamics becomes similar for large $N$. Similar phenomena are also observed for average spin dynamics $\sum_i \langle\sigma_z^i(t)\rangle / N$ when we prepare $M=N$ total excitations. In Fig.~\ref{fig4}(g) we observe a reduction in the oscillation amplitude for $N=4$ ions compared with (a) and (b), and there is a further reduction to $N=20$ as shown in Fig.~\ref{fig4}(h). However, the curves for $N=20$ and $N=32$ are again similar and show non-Markovian revivals. Note that in (h) the data points are directly connected to guide the eye, because exact diagonalization for the large Hilbert space dimensions is not feasible.

In summary, we have demonstrated quantum simulation of the JCH model using up to 32 ions and up to 32 total excitations, which amounts to $2^{77}$ dimensional Hilbert space. The conservation of excitation number in this model allows us to adjust its effective dimension, thus providing efficient verification of the quantum simulation results even for large system sizes. Similar schemes shall also work for the quantum simulation of other models where under certain parameters the effective dimension is governed by a conservation law. With this tool, we observe the change from Markovian to non-Markovian spin dynamics by tuning the spin frequencies to different locations of the phonon band. We further study the dependence of the collapse and revival signals versus the effective dimension of the Hilbert space, and find similar dynamics for large system sizes which can result from localization in the system. Our work demonstrates the trapped ion quantum simulator a powerful platform for rich properties in spin-boson coupled systems and open quantum systems with structured bosonic environments.

\begin{acknowledgments}
We thank X. Z. and L. H. for discussions. This work is supported by Tsinghua University Initiative Scientific Research Program, Beijing Academy of Quantum Information Sciences, and Frontier Science Center for Quantum Information of the Ministry of Education of China. Y.-K. W. acknowledges support from the start-up fund from Tsinghua University.
\end{acknowledgments}

%

\end{document}



\title{Supplementary Material for "Observation of Non-Markovian Spin Dynamics in a Jaynes-Cummings-Hubbard Model using a Trapped-Ion Quantum Simulator"}
\author{B.-W. Li}
\thanks{These authors contribute equally to this work}%
\affiliation{Center for Quantum Information, Institute for Interdisciplinary Information Sciences, Tsinghua University, Beijing 100084, P. R. China}
\author{Q.-X. Mei}
\thanks{These authors contribute equally to this work}%
\affiliation{Center for Quantum Information, Institute for Interdisciplinary Information Sciences, Tsinghua University, Beijing 100084, P. R. China}
\author{Y.-K. Wu}
\thanks{These authors contribute equally to this work}%
\affiliation{Center for Quantum Information, Institute for Interdisciplinary Information Sciences, Tsinghua University, Beijing 100084, P. R. China}
\author{M.-L. Cai}
\affiliation{Center for Quantum Information, Institute for Interdisciplinary Information Sciences, Tsinghua University, Beijing 100084, P. R. China}
\affiliation{HYQ Co., Ltd., Beijing, 100176, P. R. China}
\author{Y. Wang}
\affiliation{Center for Quantum Information, Institute for Interdisciplinary Information Sciences, Tsinghua University, Beijing 100084, P. R. China}
\author{L. Yao}
\affiliation{Center for Quantum Information, Institute for Interdisciplinary Information Sciences, Tsinghua University, Beijing 100084, P. R. China}
\affiliation{HYQ Co., Ltd., Beijing, 100176, P. R. China}
\author{Z.-C. Zhou}
\affiliation{Center for Quantum Information, Institute for Interdisciplinary Information Sciences, Tsinghua University, Beijing 100084, P. R. China}
\author{L.-M. Duan}
\email{lmduan@tsinghua.edu.cn}
\affiliation{Center for Quantum Information, Institute for Interdisciplinary Information Sciences, Tsinghua University, Beijing 100084, P. R. China}


\maketitle
\section*{Experimental setup}
We use a linear Paul trap to confine a chain of ${}^{171}$Yb${}^{+}$ ions. The qubit states are encoded in the hyperfine levels of ${}^{171}$Yb${}^{+}$: $|\downarrow\rangle\equiv |\mathrm{S}_{1/2}, F = 0, m_F = 0\rangle$ and $|\uparrow\rangle\equiv|\mathrm{S}_{1/2}, F = 1, m_F = 0\rangle$.
We use the routine $369.5\,$nm laser for Doppler cooling, optical pumping and qubit state detection, and use two counter-propagating $355\,$nm Raman laser beams to generate the spin-phonon coupling.
To initialize the phonon modes into the ground state, we perform multi-mode sideband cooling using the Raman beams after the Doppler cooling. Since our Raman beams have an angle of $45^\circ$ to both transverse $x$ and $y$ directions, these modes can be accessed with almost equal strength. Although we only use the $x$ modes in the experiment, we also cool the $y$ modes (which are set to be about $280\,$kHz away for 2-20 ions, and about $600\,$kHz in the 32-ion case) to suppress the undesired crosstalk. Detailed mode frequencies and ion distances for each experiment are presented in later sections.
More details about these daily operations on our setup can be found in our previous work \cite{mei2021experimental}.

\begin{figure}[!bp]
	\centering
	\includegraphics[width=0.8\linewidth]{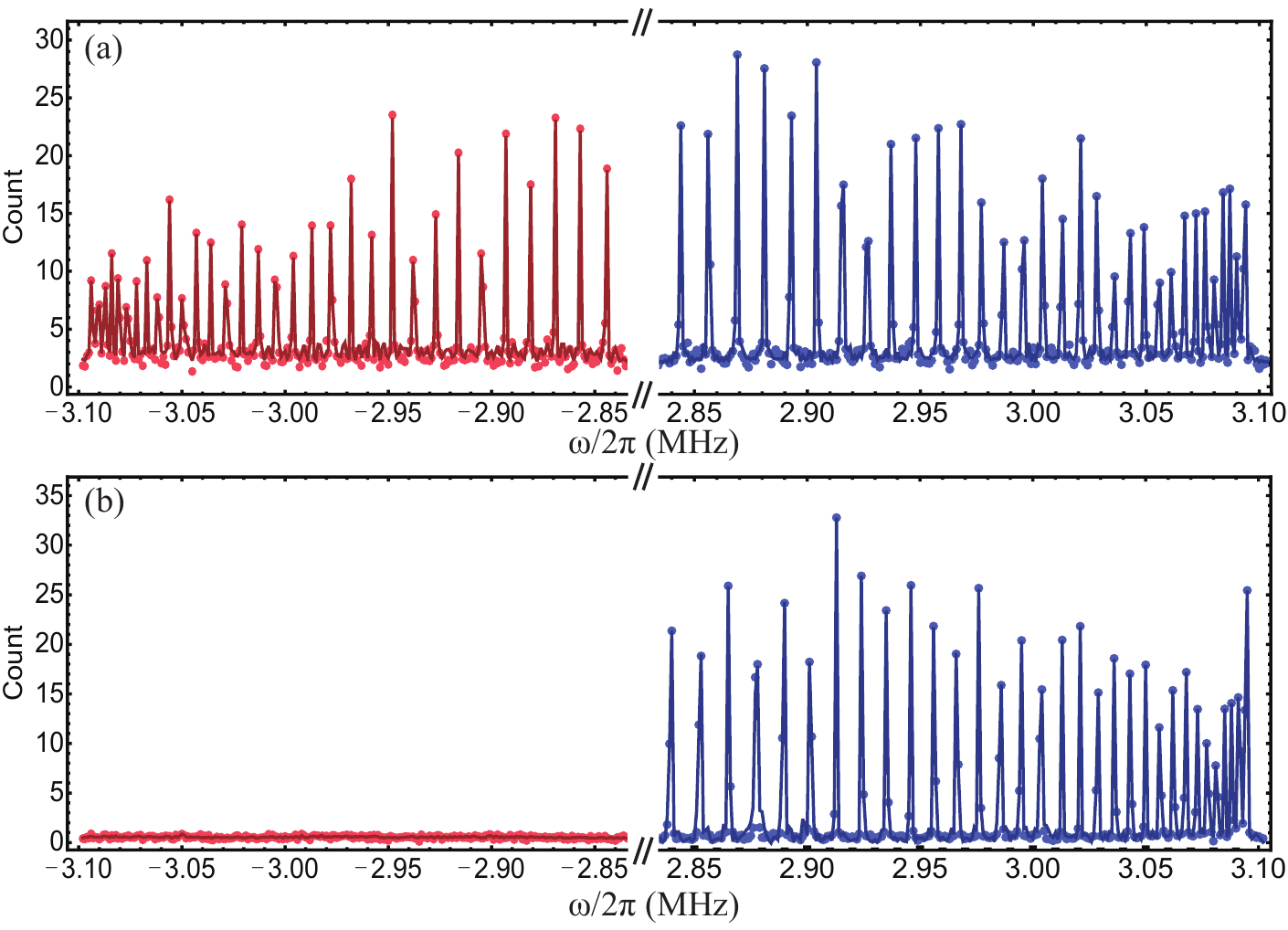}
	\caption{\label{fig:cooling}{Spectrum of the red/blue phonon sidebands after (a) $\sim1\,$ms Doppler cooling; (b) $\sim1\,$ms Doppler cooling followed by $\sim5\,$ms Raman sideband cooling. Using the ratio of the red and the blue sidebands \cite{leibfried2003quantum}, we bound the total phonon number in all the 32 modes to be below 0.97. The actual phonon number can be lower since the red sideband spectrum is affected by a nonzero background. We further fit the data by sinc functions as the solid lines, from which we extract the peaks in the spectrum with the background removed. Then we bound the total phonon number to be as low as 0.16.}}
\end{figure}

As shown in Fig.~\ref{fig:cooling}, after sideband cooling, the total phonon number in all the modes can be estimated to be about $0.16$ for the largest $N=32$ case. The actual evolution will thus be a mixture of small phonon numbers in all the modes. When we study the non-Markovian dynamics of the spins and regard the phonon modes as environments, these small excitations will not have significant effects and may explain the small deviation between the theoretical and the experimental results shown in Fig.~2 of the main text for $N=20$ and $N=32$.

The spin states can be manipulated by microwave pulses with high fidelity, but such pulses are not site-resolved and thus can only be used for global operations such as to prepare $|\uparrow\rangle^{\otimes N}$. In order to excite a selected spin state, we add a tightly focused $355\,$nm laser beam at an angle of $45^\circ$ to the quantization axis. This beam can induce fourth-order AC Stark shift \cite{PhysRevA.94.042308} on the qubit levels, which is maximized by tuning the polarization to be $\frac{1}{\sqrt{2}}\pi+\frac{1}{2}(\sigma^-+\sigma^+)$. The beam diameter is about $15\,\mu$m, capable of addressing two ions at the edge simultaneously (if only one target ion is desired, we can further deflect the laser beam outside the range of the chain). With this tightly focused beam, we use a Ramsey type experiment to prepare pure spin excitations as follows (for example if we want to excite ions $N-1$ and $N$)
\begin{enumerate}
  \item All qubits are initialized to $|\psi_0\rangle=|\downarrow\rangle^{\otimes N}$ via optical pumping.
  \item A calibrated global microwave $\pi/2$-pulse rotates all spins to the equator of the Bloch sphere $\frac{1}{2^{N/2}}\left(|\downarrow\rangle+|\uparrow\rangle\right)^{\otimes N}$.
  \item The tightly focused beam introduce fourth-order AC Stark shift to the target ions to accumulate a $\pi$ rotation around the $z$ axis, while the other ions evolve freely. We get $\frac{1}{2^{N/2}}\left(|\downarrow\rangle+|\uparrow\rangle\right)^{\otimes N-2}\otimes\left(|\downarrow\rangle-|\uparrow\rangle\right)^{\otimes 2}$
  \item A second global microwave $\pi/2$-pulse with the opposite phase rotates the spins to $|\downarrow\rangle^{\otimes N-2}\otimes |\uparrow\rangle^{\otimes 2}$
\end{enumerate}
The same sequence can also be applied to prepare one spin excitation by changing the location of the tightly-focused beam.

After simulating the dynamics of the JCH Hamiltonian, we turn off the spin-phonon coupling and use an EMCCD to measure the site-resolved ion fluorescence under $369.5\,$nm detection laser with a detection period of $450\,\mu$s \cite{mei2021experimental}. During this detection time, the phonon states can evolve, but the spins will not be affected with the coupling off.

\section*{Quantum simulation of JCH Hamiltonian}
Here we describe how the JCH model can be realized in our setup. Following our previous work \cite{mei2021experimental}, the Hamiltonian for the motional states can be expressed as
\begin{equation}
H_m = \frac{1}{2} \sum_i \left(m\omega_x^2 - \frac{e^2}{4\pi\epsilon_0} \sum_{j\ne i} \frac{1}{z_{ij}^3} \right) x_i^2 + \sum_{i<j} \frac{e^2}{4\pi\epsilon_0} \frac{1}{z_{ij}^3} x_i x_j + \sum_i \frac{p_i^2}{2m},
\end{equation}
where $z_{ij}\equiv|z_i-z_j|$ is the equilibrium distance between two ions $i$ and $j$, $x_i$ and $p_i$ the transverse position and momentum of the ion $i$, and $\omega_x$ the transverse trap frequency.
We define local trap frequency $\omega_i \equiv \sqrt{\omega_x^2 - (e^2/4\pi\epsilon_0 m)\sum_{j\ne i}1/z_{ij}^3}$ and quantize the local oscillation as $x_i=\sqrt{\hbar/2m\omega_i}(a_i+a_i^\dag)$ and $p_i=i\sqrt{\hbar m \omega_i / 2} (a_i^\dag - a_i)$. Assuming $e^2/4\pi\epsilon_0 m \omega_x^2 z_{ij}^3 \ll 1$ and performing rotating wave approximation, we get \cite{mei2021experimental}
\begin{equation}
H_m = \sum_i \tilde{\omega}_i a_i^\dag a_i + \sum_{i<j} \tilde{t}_{ij} (a_i^\dag a_j + a_j^\dag a_i), \label{eq:motional_RWA}
\end{equation}
where
\begin{equation}
\omega_i = \sqrt{\omega_x^2 - \frac{e^2}{4\pi\epsilon_0 m}\sum_{j\ne i}\frac{1}{z_{ij}^3}},
\end{equation}
\begin{equation}
t_{ij}=\frac{e^2}{8\pi\epsilon_0 m \sqrt{\omega_i\omega_j} z_{ij}^3},
\end{equation}
and
\begin{equation}
\tilde{\omega}_i = \omega_i - \frac{1}{2\omega_x}\sum_{j\ne i}t_{ij}^2,
\end{equation}
\begin{equation}
\tilde{t}_{ij} = t_{ij} - \frac{1}{2\omega_x} \sum_{k\ne i,j} t_{ik} t_{jk}.
\end{equation}

Now we add the spin Hamiltonian $H_s=\sum_i(\omega_{01}/2)\sigma_z^i$ and the coupling Hamiltonian $H_{r}=\sum_i\Omega_i\cos(k_{r}x_i-\omega_{r}t+\phi_{r})\sigma_x^i$ for a laser beam near the red motional sideband with $\omega_r=\omega_{01}-\omega_x-\Delta$. In an interaction picture with $H_0=\sum_i [(\omega_{01}-\Delta)/2]\sigma_z^i + \sum_i \omega_x a_i^\dag a_i$, we get
\begin{equation}
H_I = \sum_i \frac{\Delta}{2}\sigma_z^i + \sum_i (\tilde{\omega}_i - \omega_x) a_i^\dag a_i + \sum_{i<j} \tilde{t}_{ij} (a_i^\dag a_j + a_j^\dag a_i) + \sum_i \frac{\eta_i\Omega_i}{2} \left( \sigma_+^i a_i + a_i^\dag \sigma_-^i\right).
\end{equation}
We can thus identify each term with that in Eq.~(1) of the main text.

\section*{Calibrating experimental parameters}
\begin{figure}[!tbp]
	\centering
	\includegraphics[width=\linewidth]{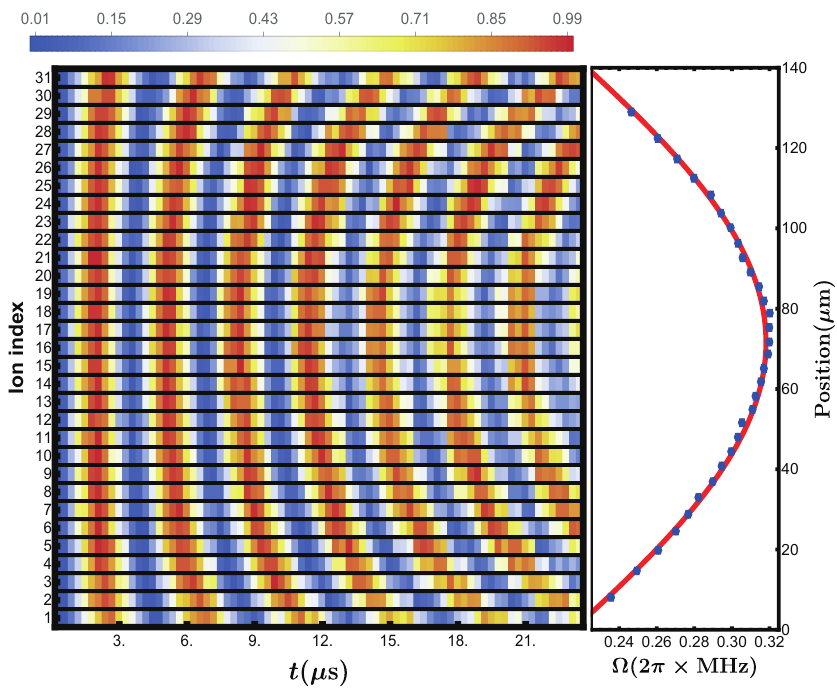}
	\caption{\label{figs_Gaussian}Left: Carrier Rabi oscillation of $N=31$ ions. Right: Position-dependent Rabi frequencies are fitted by a Gaussian function. }
\end{figure}

To calibrate $\omega_i$ and $t_{ij}$, we scan the detuning of the global Raman beams around the red motional sideband to get the spectrum of the collective phonon modes. The frequencies of the $N$ collective modes allow us to fit the inter-ion spacings \cite{mei2021experimental}, which in turn give us the inter-site hopping rates and the local trap frequencies.
Our global Raman beams to generate the spin-phonon coupling have a Gaussian waist size (lengths of the major axis and the minor axis where the intensity drops to $1/e^2$) of $300\times 20\,\mu$m, so the coupling strength is not uniform over the long ion chain with a length of about $100\,\mu$m. We calibrate this position-dependent driving strength by tuning the Raman laser to resonance with the carrier transition of the ions and measure the Rabi oscillation, as shown in the left panel of Fig.~\ref{figs_Gaussian}. This allows us to extract the local driving strength $\Omega_i$ at the position of $z_i$, from which we fit a Gaussian function $\Omega_0 e^{-2z^2/\sigma^2}$ with $\sigma=162\,\mu$m. Now by tuning the laser intensity, we can set the local spin-phonon coupling strength $g_i=\eta_i\Omega_i/2$ for the central ion to the desired value, and then the coupling strength for the other ions can be computed correspondingly. As for the calibration of $\Delta_i$ (which is subjected to AC Stark shift), note that if we place a single ion at different positions $z_i$ under the given Raman laser beams, then $\Delta_i$ is just the Raman laser detuning from the center-of-mass mode of the red phonon sideband. In principle we can move the ion along the axial direction to calibrate $\Delta_i$ at different $z_i$, but given the above distribution of the laser amplitudes, here we can similarly fit the AC Stark shift over the long ion chain as $D_0 e^{-4z^2/\sigma^2}$ where $D_0$ is the AC Stark shift of the central ion.  With this function, $\Delta_i$ for different ions can be computed accordingly.

\section*{Experimental parameters}
Here we specify the experimental parameters used in Fig.~2, Fig.~3 and Fig~.4 of the main text. In particular, we present the measured collective phonon mode frequencies and the fitted ion distances. With $\Delta$ and $g$ set in the main text, other parameters can be computed using the above formulas.

For $N = 2$ ions in Fig.~2, measured collective phonon mode frequencies are $2\pi\times(2.666,\, 2.718)\,$MHz, and the fitted ion distance is $\Delta z = 5.280 \,\mu$m. These correspond to local phonon frequencies $\omega_i = 2\pi\times(-25.96,\,-25.96)\,$kHz in the interaction picture.

For $N = 4$ ions in Fig.~3, measured collective phonon mode frequencies are $2\pi\times(2.692,\, 2.714,\,2.732,\,2.744)\,$MHz (thus the phonon band in the interaction picture ranges from $-2\pi\times 52\,$kHz to zero), and the fitted ion distances are $\Delta z = (6.602,\,6.104,\, 6.602)\,\mu$m. These correspond to local phonon frequencies $\omega_i = 2\pi\times(-15.45,\,-31.61,\,-31.61,\,-15.45)\,$kHz in the interaction picture.

For $N = 4$ ions in Fig.~4, measured collective phonon mode frequencies are $2\pi\times(2.689,\,2.711,\, 2.729,\,2.741)\,$MHz, and the fitted ion distances are $\Delta z = (6.588,\,6.102,\,6.588)\,\mu$m. This corresponds to local phonon frequencies $\omega_i = 2\pi\times(-15.56,\,-31.75,\,-31.75,\,-15.56)\,$kHz in the interaction picture.

For $N = 8$ ions in Fig.~2, measured collective phonon mode frequencies are $2\pi\times(2.689,\,2.703,\,2.716,\,2.727,\,2.737,\,2.745,\\2.751,\,2.756)\,$MHz, and the fitted ion distances are $\Delta z = (7.655,\,6.496,\,6.030,\,5.940,\,6.030,\,6.496,\,7.655)\,\mu$m. These correspond to local phonon frequencies $\omega_i = 2\pi\times(-10.55,\,-25.04,\,-35.46,\,-40.58,\,-40.58,\,-35.46,\,-25.04,\,-10.55)\,$kHz in the interaction picture.

For $N = 20$ ions in Fig.~2, Fig.~3 and Fig.~4, collective phonon mode frequencies are $2\pi\times(2.575,\,2.589,\,2.602,\,2.615,\,2.627,\\2.639,\,2.650,\,2.661,\,2.671,\,2.681,\, 2.690,\,2.698,\,2.706,\,2.713,\,2.720,\,2.726,\,2.731,\,2.735,\,2.739,\,2.742)\,$MHz, and the fitted ion distances are $\Delta z = (7.290,\,6.236,\,5.594,\,5.210,\,4.960,\,4.792,\,4.672,\, 4.590,\,4.561,\,4.516,\,4.561,\,4.590,\,4.672,\,4.792,\\4.960,\,5.210,\,5.594,\,6.236,\,7.290)\,\mu$m. These correspond to local phonon frequencies $\omega_i = 2\pi\times(-12.55,\,-29.37,\,-43.98,\\
-57.54,\,-69.10,\,-78.58,\,-86.18,\,-92.06,\,-95.63,\,-97.97,\,-97.97,\,-95.63,\,-92.06,\,-86.18,\,-78.58,\,-69.10,\\
-57.54,\,-43.98,\,-29.37,\,-12.55)\,$kHz in the interaction picture.

For N = 32 ions in Fig.~2, Fig.~3 and Fig.~4, collective phonon mode frequencies are $2\pi\times(2.603,\,2.616,\,2.630,\,2.643,\,2.656,\\
2.668,\,2.680,\,2.692,\,2.704,\,2.715,\,2.726,\, 2.737,\,2.747,\,2.757,\,2.767,\,2.776,\,2.785,\,2.794,\,2.802,\,2.810,\,2.817,\,2.824,\\2.831,\, 2.838,\,2.843,\,2.849,\,2.854,\,2.858,\,2.862,\,2.865,\,2.868,\,2.872)\,$MHz, and the fitted ion distances are $\Delta z = (6.661,\\
5.859,\,5.328,\,4.920,\,4.668,\,4.471,\,4.312,\,4.195,\,4.099,\,4.049,\,3.968,\,3.929,\, 3.892,\,3.868,\,3.857,\,3.867,\,3.857,\,3.868,\\
3.892,\,3.929,\,3.968,\,4.049,\,4.099,\,4.195,\, 4.312,\,4.471,\,4.668,\,4.920,\,5.328,\,5.859,\,6.661)\,\mu$m. These correspond to local phonon frequencies $\omega_i = 2\pi\times(-15.59,\,-34.82,\,-49.71,\,-64.89,\,-79.19,\,-91.83,\,-103.73,\,-114.42,\,-123.72,\,-130.87,\\
-137.71,\,-144.13,\,-148.63,\,-152.25,\,-154.39,\,-154.63,\,-154.63,\,-154.39,\,-152.25,\,-148.63,\,-144.13,\,-137.71,\\
-130.87,\,-123.72,\,-114.42,\,-103.73,\,-91.83,\,-79.19,\,-64.89,\,-49.71,\,-34.81,\,-15.59)\,$kHz in the interaction picture.

\bibliographystyle{apsrev4-1-title}
\bibliography{reference}